# Hybrid quantum-classical physics-informed neural networks for solving nonlinear PDEs: when and where hybridization is effective?

Kaveh Zabihi[1], Hamid Montazeri[2,3], Akke S.J. Suiker[1]

*[1]Chair of Applied Mechanics, Department of the Built Environment, Eindhoven University of Technology, PO Box 513, 5600 MB Eindhoven, The Netherlands*
*[2]Power & Flow Group, Department of Mechanical Engineering, Eindhoven University of Technology, PO Box 513, 5600 MB Eindhoven, The Netherlands*
*[3]Center for Quantum Materials and Technologies Eindhoven (QT/e), P.O. Box 513, 5600 MB Eindhoven, The Netherlands*

**Abstract:** Physics-informed neural networks (PINNs) often struggle on nonlinear partial differential equations (PDEs) with sharp gradients, stiff dynamics, high-frequency content, or multiscale structure. Such limitations, rooted in spectral bias, ill-conditioned optimization, and unstable convergence, restrict PINN accuracy in the regimes where advanced solvers are most needed. In this work, we develop a hybrid quantum-classical physics-informed neural network (HQPINN) that integrates a classical neural-network backbone with a parameterized quantum circuit (PQC) to enrich the solution representation. The framework is benchmarked against a classical PINN on three representative nonlinear PDEs, Burgers' equation, the Allen–Cahn equation, and the Korteweg–de Vries (KdV) equation. The framework is further examined through a systematic sensitivity analysis of qubit count, circuit depth, PQC placement, collocation density, and classical-network width. Across all benchmarks, HQPINNs exhibit smoother training dynamics, reduced loss oscillations, and improved final accuracy, with the largest gains occurring in stiff and multiscale regimes. Relative $L^2$ error decreases by about fourfold for Burgers' equation and fivefold for the Allen–Cahn equation, while improvements for the KdV equation are more moderate. Overall, the results demonstrate that carefully co-designed hybrid quantum–classical architectures can mitigate key limitations of classical PINNs and provide practical design guidance for near-term quantum-enhanced PDE solvers.

## 1 Introduction

Nonlinear partial differential equations (PDEs) play a central role in the mathematical modeling of complex physical phenomena, including fluid dynamics, phase-transition processes, and materials science. However, many practically relevant regimes remain difficult to solve accurately, particularly when the solution exhibits sharp gradients, stiff interface dynamics, multiscale interactions, or rapidly varying oscillatory structures. Such behaviors introduce strong nonlinear coupling, localized features, and competing spatial and temporal scales that simultaneously challenge numerical accuracy, stability, and efficiency. As a result, the robust solution of nonlinear PDEs in these demanding regimes remains a major challenge in computational science and engineering.

In recent years, physics-informed neural networks (PINNs) have emerged as a promising paradigm at the interface of machine learning and scientific computing (Michaloglou et al., 2025; Raissi et al., 2019). In this approach, the governing PDE residual together with the associated initial and boundary conditions are embedded directly into the loss function, thereby reformulating PDE solving as a physics-constrained optimization problem. By penalizing deviations from the governing equations during training, PINNs can approximate continuous solution fields over the domain without requiring a mesh or an explicit discrete solver. This mesh-free formulation, combined with the ability to treat forward and inverse problems within unified framework and to incorporate additional data or constraints naturally, has made PINNs an attractive alternative to conventional discretization-based numerical methods (Sahli Costabal et al., 2024).

Despite their promise, classical PINNs remain challenged when applied to PDEs that exhibit strong nonlinearity, high-frequency content, or stiff reaction–diffusion dynamics. A well-known limitation is spectral bias, whereby neural networks tend to learn low-frequency components first and therefore struggle to accurately represent sharp gradients, localized structures, or rapidly oscillatory features. As a result, classical PINNs often require large network architectures, dense collocation sampling, or prolonged training to resolve shocks, metastable interfaces, or multiscale solution behavior (Tancik et al., 2020). Moreover, the associated optimization problem can become stiff and poorly conditioned due to imbalanced gradients among the PDE-residual, initial-condition, and boundary-condition terms, leading to slow or unstable convergence. These limitations become especially severe in the nonlinear PDE regimes considered in this work, where sharp, high-frequency, or multiscale dynamics play a dominant role (Cuomo et al., 2022).

In this context, parameterized quantum circuits (PQCs) have attracted growing attention as expressive function approximators in scientific computing (Cerezo et al., 2021; de Keijzer et al., 2025; Acampora et al., 2025). PQCs exhibit high expressive capacity in high-dimensional Hilbert spaces, suggesting that quantum layers may enrich solution representations (Larocca et al., 2023). Such representations are particularly relevant for PDE regimes described above, where classical PINNs often struggle because of spectral bias and optimization challenges. PQCs therefore provide a natural motivation for exploring hybrid physics-informed learning frameworks that combine quantum representational flexibility with classical optimization and physics-based constraint enforcement.

These considerations motivate hybrid quantum-classical physics-informed neural networks (HQPINNs), in which parameterized quantum circuits are embedded within classical PINN architectures (Markidis, 2022). Rather than replacing the classical model, the hybrid architecture augments it with a quantum layer that enriches the learned solution representation. In this framework, the classical neural backbone extracts a structured latent representation of the input coordinates, while the PQC operates on this latent space to introduce additional nonlinear and oscillatory structure. HQPINNs therefore provide a principled way to combine the optimization robustness of classical neural networks with the representational flexibility of quantum circuits, particularly for challenging nonlinear PDEs.

A growing body of work has explored the integration of quantum circuits into physics-informed learning frameworks for solving differential equations, as summarized in Table 1. Across these studies, both pure quantum and hybrid quantum-classical architectures have been investigated using qubit-based and continuous-variable implementations, with applications ranging from Burgers' and Poisson equations to Euler, Navier–Stokes, Helmholtz, and Schrödinger-type equations. Collectively, these studies establish the feasibility of incorporating quantum models into PINN formulations and suggest that compact quantum or hybrid architectures can achieve competitive accuracy in selected settings. At the same time, the existing literature remains highly heterogeneous in its design choices, problem settings, and evaluation criteria. Circuit sizes are typically small, benchmark PDEs vary widely, and architectural choices such as quantum-layer placement, circuit depth, and interaction with classical components differ substantially across studies. As a result, reported performance gains are difficult to compare directly, and it remains unclear whether observed improvements arise from the quantum component itself, favorable problem structure, or specific architectural and training choices. In particular, the literature still lacks a unified framework for assessing how quantum resources, classical-network capacity, and training design jointly influence performance across distinct PDE regimes.

To address these gaps, this study develops a hybrid quantum-classical physics-informed neural network (HQPINN) framework and systematically benchmarks its performance across distinct nonlinear PDE regimes. The proposed model is compared directly against a classical PINN on three representative equations: Burgers' equation, the Allen–Cahn equation, and the Korteweg–de Vries (KdV) equation. These equations are chosen deliberately to probe complementary numerical challenges. Burgers' equation tests the ability to resolve sharp gradients and shock-forming behavior, the Allen–Cahn equation probes stiff reaction–diffusion dynamics and metastable interfacial structure, and the KdV equation assesses performance in the presence of higher-order dispersive effects. Together, they provide a diverse and well-controlled testbed for assessing how

**Table 1.** Literature review of variational quantum and continuous-variable approaches for physics-informed neural network.

| Reference | Simulator | Circuit Arch. (No. qubits/qumodes) | Sens. Analysis | NN Type | PDE Type (Dimension) |
|---|---|---|---|---|---|
| Yew Leong et al., 2025 | PL | PQC (3) | Arch, L | Hybrid | Euler (1D); Transonic airfoil flow (2D) |
| Sedykh et al., 2024 | QMware | VQC (3) | Qb, L | Hybrid | NS (3D) |
| Farea et al., 2025 | PL + SF | DV (2–5) | Qb, Topo, Emb | Hybrid | Helmholtz (1D); NS (2D); Wave (1D); KG (1D) |
| Trahan et al., 2024 | PL | PQC (2–5) | Qb, L, Noise, Nodes | Pure/Hybrid | Spring-mass (1D); Poisson (2D); Burgers (1D) |
| Panichi et al., 2026 | SF | CVQC (2*) | Noise | Pure | Poisson (1D); Heat (1D) |
| Dehaghani et al., 2024 | SF | CVQC (2*) | L, Noise | Hybrid | Quantum optimal ctrl; Pontryagin min. principle |
| Vadyala & Betgeri, 2023 | PyQu | CVQC (1–2*) | L, Nn, LR | Pure | Schrödinger (1D) |
| Xiao et al., 2024 | PL | PQC (3–5) | L, Arch, Emb | Pure | Potential flow (2D); Burgers (2D), Energy (2D), Euler (2D) |
| Kyriienko et al., 2021 | Julia/Yao.jl | DQC (6) | L, Arch | Pure | NS (quasi-1D) |
| Markidis, 2022 | SF | PQC (1–2*) | L, Opt, Res, BS | Pure | Poisson (1D) |
| Berger et al., 2025 | PL | PQC (2–8) | Qb, L, Arch | Hybrid | ODE, Poisson (2D); Burgers (1D); NS (2D) |
| Setty et al., 2025 | PL | PQC (3–8) | Arch, W, Obs, LR | Self-adaptive quantum | Riccati; ODE, Poisson (2D) |

PL = PennyLane; SF = Strawberry Fields; PQC = parameterized quantum circuit; VQC = variational quantum circuit; DQC = differentiable quantum circuit; QNN = quantum neural network; CV = continuous variable; CVQC = continuous-variable quantum circuit; QPINN = quantum PINN; Qb = number of qubits; L = number of quantum/circuit layers; Nn = number of neurons; LR = learning rate; Obs = observables; Opt = optimizer; Arch = architecture; Emb = embedding/encoding strategy; Topo = topology; Nodes = collocation/training nodes; W = loss/self-adaptive weights; NS = Navier–Stokes; KG = Klein–Gordon; Pure = no classical hidden layers; Hybrid = classical and quantum components.

*For continuous-variable quantum circuits, the number refers to qumodes rather than qubits.

HQPINNs behave across qualitatively different classes of nonlinear solution behavior.

In addition to this baseline comparison, the study conducts a comprehensive sensitivity analysis of qubit count, variational depth, quantum-layer placement, sampling density, and classical-network width. This design enables a rigorous, systematic, and comparative assessment of HQPINNs, providing insight into how hybrid quantum–classical architectures behave across PDE regimes with varying spectral complexity, multiscale structure, and stiffness. The main contributions of this work are threefold. First, it develops an HQPINN framework for challenging nonlinear PDE regimes in which classical PINNs often struggle. Second, it establishes a systematic benchmark against classical PINNs across three representative PDE classes with distinct numerical characteristics. Third, it presents a comprehensive sensitivity analysis showing how quantum-resource allocation, hybrid architectural design, and sampling conditions shape model performance. Together, these contributions provide both empirical insight and practical design guidance for hybrid quantum–classical PDE solvers.

The outline of the paper is as follows. Section 2 presents the methodology, followed by a detailed description of the proposed HQPINN in Section 3. Section 4 describes the benchmark PDEs, implementation framework, and computational settings. Section 5 reports the baseline and sensitivity analysis results. Finally, Sections 6 and 7 present the discussion and conclusions, respectively.

## 2 Methodology

A hybrid quantum-classical physics-informed neural network (HQPINN) framework is developed for solving nonlinear PDEs. The proposed HQPINN architecture integrates a classical feed-forward neural-network backbone with a parameterized quantum circuit (PQC) into a single end-to-end differentiable model. All model components are trained jointly using the same physics-informed loss formulation. The detailed HQPINN architecture, quantum-circuit ansatz, and data-encoding strategy are described in Section 3. The framework is evaluated on three benchmark PDEs: Burgers' equation, the Allen–Cahn equation, and the KdV equation. Together, these equations represent shock-dominated dynamics, stiff reaction-diffusion behavior with metastable interfaces, and higher-order dispersive effects. A unified evaluation approach is employed across all benchmarks, using identical error metrics, convergence analysis, and training schedules. This controlled setup ensures consistency across experiments while allowing each PDE to probe distinct computational and optimization challenges. Mathematical details of the benchmark equations are provided in Section 4.1. To enable direct and fair comparison, classical PINNs are implemented as baseline models. These baselines use the same classical neural backbone, input normalization, collocation-point sampling strategy, optimizer sequence, and training schedule as their HQPINN counterparts. As a result, observed performance differences can be attributed directly to the inclusion and configuration of the quantum component, rather than to confounding implementation choices.

A central component of this study is a systematic sensitivity analysis examining how HQPINN performance depends on key architectural and training parameters. Five components are examined independently: (i) the number of qubits, (ii) the number of variational layers, (iii) the placement of the PQC within the hybrid architecture, (iv) the number of training samples, and (v) the number of neurons in the classical network. In each study, a single factor is varied while others are held fixed, enabling direct assessment of its impact on accuracy and training behavior. This enables a controlled investigation of classical–quantum co-design effects across different PDE regimes.

All experiments are performed using classical quantum-circuit simulators. This choice helps isolate the behavior of the hybrid architecture from hardware-specific factors such as gate noise, readout errors, crosstalk, and execution latency, which are intentionally excluded from the present study.

## 3 HQPINNS: model development

### 3.1 Overview of the HQPINN architecture

The HQPINN model maps the spatiotemporal coordinates $(x, t)$ to the scalar solution field $u(x, t)$ through a hybrid classical-quantum architecture, as illustrated in Fig. 1. The model consists of three main stages. First, the input coordinates are processed by a classical feed-forward neural network, which transforms the low-dimensional physical inputs into a compact latent representation. This classical backbone serves to extract structured, problem-adapted features from the input domain. Second, the latent representation is passed to a PQC, where it is encoded into a quantum state and transformed by a variational quantum ansatz. The quantum circuit acts as a nonlinear feature transformation that enriches the learned representation with additional oscillatory structure. Third, expectation values measured from the quantum circuit are mapped through a final classical layer to produce the scalar network output $u(x, t)$. Since both the classical neural network and the PQC are differentiable, gradients of the network output with respect to the input coordinates and all trainable parameters can be computed using automatic differentiation. This enables all classical and quantum parameters to be optimized

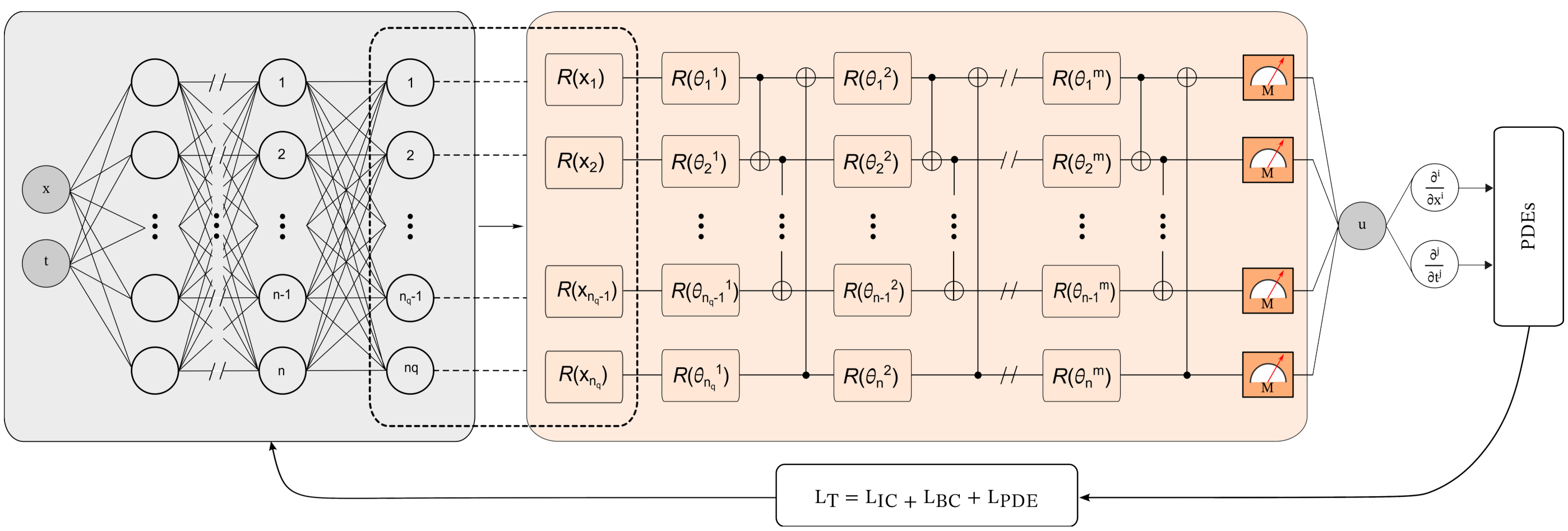

**Fig. 1** Schematic of the proposed hybrid quantum-classical physics-informed neural network (HQPINN) architecture. The spatiotemporal input $(x, t)$ is first processed by a classical feed-forward neural network (depicted in gray) to produce a compact latent representation, which is then encoded into an $n_q$-qubit parameterized quantum circuit (PQC) using angle embedding, followed by $m$ variational layers composed of single-qubit rotations and a ring CNOT entangling gates (depicted in orange). Pauli-Z expectation values measured from the PQC are subsequently mapped through a final classical dense layer to generate the scalar network prediction $u(x, y)$. Automatic differentiation is used to compute spatial and temporal derivatives required by the composite physics-informed loss $\mathcal{L}_T = \mathcal{L}_{\mathrm{IC}} + \mathcal{L}_{\mathrm{BC}} + \mathcal{L}_{\mathrm{PDE}}$, and all classical and quantum parameters are optimized jointly within an end-to-end differentiable framework.

jointly within a unified physics-informed training framework.

### 3.2 Classical backbone

The classical component of the HQPINN consists of a fully connected feed-forward neural network with 6 hidden layers, each comprising 40 neurons and using hyperbolic tangent activation functions. All network weights are initialized using the Glorot normal initialization scheme. To improve numerical stability during training, the input data are normalized before being passed through the network. The output of the final classical hidden layer provides a structured latent representation that is transferred to the quantum encoding stage.

### 3.3 Classical-to-quantum encoding

The classical backbone maps the spatiotemporal input coordinates to a compact latent representation that serves as the interface to the quantum subnetwork. Let $\boldsymbol{h} \in \mathbb{R}^{40}$ denote the output of the penultimate classical layer. The final classical dense layer preceding the parameterized quantum circuit applies a hyperbolic tangent activation function and projects $\boldsymbol{h}$ onto an $n_q$-dimensional latent vector:

$$\mathbf{z} = \left(z_1, z_2, \dots, z_{n_q}\right) \in [-1,1]^{n_q} \tag{1}$$

where $n_q$ denotes the number of qubits in the quantum circuit. Each latent component is defined as:

$$z_i = \tanh(\mathrm{w}_i^{\top} h + b_i), \quad i = 1, \dots, n_q \tag{2}$$

with $\mathbf{w}_i^{\top} \in \mathbb{R}^{40}$ and $b_i \in \mathbb{R}$ representing the trainable weight vector and bias associated with the $i$-th latent unit. The hyperbolic tangent activation ensures that the latent variables remain bounded, which is well suited for subsequent angle-based quantum encoding.

The latent vector $\mathbf{z}$ is encoded into the quantum state of an $n_q$-qubit register using angle embedding. Each qubit is initialized in the computational ground state $|0\rangle$, and a single-qubit rotation about the x-axis ($R_X$) is applied (Bergholm et al., 2022; Schuld et al., 2021) with the rotation angle given by the corresponding latent component $z_i$:

$$R_X(z_i) = \exp\left(-\frac{i}{2} z_i\, \sigma_x\right) = \begin{pmatrix} \cos(\frac{z_i}{2}) & -i\sin(\frac{z_i}{2}) \\ -i\sin(\frac{z_i}{2}) & \cos(\frac{z_i}{2}) \end{pmatrix} \tag{3}$$

where $\hat{\sigma}_x$ denotes the Pauli-X operator. The resulting state of the $i$-th qubit is:

$$|\psi_i(z_i)\rangle = R_X(z_i)|0\rangle = \cos\left(\frac{z_i}{2}\right)\Big|0\rangle - i\sin\left(\frac{z_i}{2}\right)\Big|1\rangle \tag{4}$$

The full quantum state after encoding is given by the tensor product over all qubits:

$$\begin{aligned} |\Psi_{enc}(z)\rangle &= \bigotimes_{i=1}^{n_q} R_X(z_i)\ |0\rangle \\ &= \bigotimes_{i=1}^{n_q} \left[\cos\left(\frac{z_i}{2}\right)\Big|0\rangle - i\sin\left(\frac{z_i}{2}\right)\Big|1\rangle\right] \end{aligned} \tag{5}$$

In the present framework, the bounded latent vector produced by the final tanh layer is passed directly to the angle embedding layer without additional rescaling.

### 3.4 Variational quantum circuit and measurement ansatz

Following the angle-encoding stage, the encoded quantum state is processed by a parameterized variational quantum circuit (PQC), which serves as the trainable quantum feature transformation within the HQPINN framework. The PQC is implemented using $m$ variational layers, each consisting of trainable single-qubit rotations, $R_X$, followed by a ring entangling block of CNOT gates. Specifically, in layer $l = 1, \dots, m$, a rotation $R_X(\theta_{l,i})$ is applied to each qubit $i$, followed by a ring-entanglement pattern implemented using CNOT gates between adjacent qubits (Bergholm et al., 2022). The overall variational operator is therefore written as:

$$U(\boldsymbol{\theta}) = \prod_{l=1}^{m} \left[ \mathrm{CNOT}_{\mathrm{ring}} \cdot \bigotimes_{i=1}^{n_q} R_X(\theta_{m,i}) \right] \tag{6}$$

where $\boldsymbol{\theta} \in \mathbb{R}^{m \times n_q}$ denotes the set of trainable quantum parameters. The resulting output state is:

$$|\Psi_{\mathrm{out}}(\mathbf{z}, \boldsymbol{\theta})\rangle = U(\boldsymbol{\theta})\, |\Psi_{\mathrm{enc}}(\mathbf{z})\rangle \tag{7}$$

Quantum features are extracted by measuring the expectation value of the Pauli-Z operator on each qubit:

$$o_i(\mathbf{z}, \boldsymbol{\theta}) = \langle \Psi_{\mathrm{out}} | \, \sigma_z^{(i)} \, | \Psi_{\mathrm{out}} \rangle \in [-1,1], \quad i = 1, \dots, n_q \tag{8}$$

The measurements produce a feature vector $\mathbf{o} = (o_1, \dots, o_{n_q}) \in [-1,1]^{n_q}$, which is passed to a final classical dense layer to generate the scalar network prediction $u(x,t)$. Since both the variational circuit and measurement operations are differentiable, gradients with respect to the quantum parameters $\boldsymbol{\theta}$ can be propagated as part of the end-to-end hybrid training procedure.

### 3.5 Physics-informed loss and training

The predicted solution $u(x,t)$ is trained using the standard physics-informed formulation in which the total loss function ($\mathcal{L}_T$) combines initial-condition ($\mathcal{L}_{\mathrm{IC}}$), boundary-condition ($\mathcal{L}_{\mathrm{BC}}$), and PDE-residual ($\mathcal{L}_{\mathrm{PDE}}$) terms:

$$\mathcal{L}_T = \mathcal{L}_{\mathrm{IC}} + \mathcal{L}_{\mathrm{BC}} + \mathcal{L}_{\mathrm{PDE}} \tag{9}$$

The initial-condition and boundary-condition losses are defined as mean-squared-error penalties between predicted and prescribed values at the corresponding training points. The PDE-residual term is computed by evaluating the governing equation at interior collocation points, where all required spatial and temporal derivatives of $u(x,t)$ are obtained through automatic differentiation. Because the HQPINN architecture is fully differentiable, derivative information propagates through both the classical neural-network layers and the parameterized quantum circuit. This enables all classical weights and quantum circuit parameters to be optimized jointly under a unified physics-informed objective. No separate training stages or auxiliary surrogate losses are introduced.

### 3.6 Implementation details

The HQPINN framework is implemented in Python 3.11.1 using TensorFlow 2.13.0 and Keras 2.13.1 for the classical network, and PennyLane 0.40.0 for the parameterized quantum-circuit implementation. All classical and quantum components are trained within a unified automatic-differentiation workflow. All experiments are conducted using classical quantum-circuit simulators. Consequently, the reported results correspond to noiseless quantum simulations and do not include hardware-specific effects such as gate noise, readout errors, crosstalk, or execution latency. This design choice isolates the algorithmic behavior of the hybrid quantum-classical architecture from hardware constraints and enables a controlled assessment of HQPINN performance. Unless stated otherwise, identical training settings are used across all experiments to ensure consistency and reproducibility. This implementation setup allows the proposed HQPINN framework to be reproduced using standard open-source software tools without reliance on specialized quantum hardware.

## 4 HQPINNS: application to PDEs

### 4.1 Benchmark PDEs

To evaluate the performance and generality of the proposed HQPINN framework, three representative one-dimensional nonlinear PDEs are considered: Burgers' equation, the Allen–Cahn equation, and the KdV equation. These equations are selected because they span distinct classes of nonlinear behavior that are known to challenge conventional PINNs including shock-like steepening, stiff reaction–diffusion dynamics, and dispersive wave propagation.. Together, they provide a diverse and methodologically meaningful testbed for assessing the accuracy, robustness, and problem-dependent behavior of the hybrid quantum-classical framework.

- *Burgers' equation:* The viscous Burgers' equation is a standard benchmark for nonlinear transport problems, combining convection and diffusion and exhibiting shock-like structures for small viscosity values. The governing equation is:

$$\frac{\partial u}{\partial t} + u\frac{\partial u}{\partial x} = \nu\frac{\partial^2 u}{\partial x^2} \quad (10)$$

where $u(x, t)$ is the solution field and $\nu$ is the viscosity coefficient. In this study, $\nu = 0.01/\pi$ is used, resulting in steep gradients and near-shock features that test the ability of the HQPINN to resolve nonlinear advection-diffusion behavior.

- *Allen-Cahn equation:* The Allen–Cahn equation is a prototypical reaction–diffusion model describing phase separation and metastable interface dynamics. It is given by:

$$\frac{\partial \phi}{\partial t} = D\frac{\partial^2 \phi}{\partial x^2} - f'(\phi) \quad (11)$$

where $\phi(x, t)$ is the order parameter, $D$ is the diffusion coefficient that controls the interface thickness between phases, and $f'(\phi)$ is the reaction term, typically defined as the derivative of a double-well potential $f(\phi)$. In the present study, $D = 0.0001$ and $f'(\phi) = 5\phi - 5\phi^3$. This formulation captures sharp interfaces, stiff temporal dynamics, and long-lived metastable states, providing a stringent test of the model's ability to handle nonlinear reaction-diffusion systems.

- *Korteweg–de Vries (KdV) equation:* The KdV equation governs the unidirectional propagation of nonlinear dispersive waves, including solitons. A general form may be written as:

$$\frac{\partial \psi}{\partial t} + \alpha\psi\frac{\partial \psi}{\partial x} + \mu\frac{\partial^3 \psi}{\partial x^3} = 0 \quad (12)$$

where $\psi(x, t)$ represents the wave profile, $\alpha$ is the coefficient associated with nonlinear wave steepening, and $\mu$ is the dispersion coefficient controlling the strength of dispersive effects. In the present study, $\alpha = 1$ and $\mu = 0.0025$ is used. This yields dispersive wave structures that test the ability of the HQPINN to capture higher-order derivatives and complex solution behavior.

## 4.2 Implementation framework

### *4.2.1 Data preprocessing*

Reference solutions for the three benchmark PDEs, Burgers' equation, the Korteweg–de Vries (KdV) equation, and the Allen–Cahn equation are adopted from the numerical benchmark datasets provided by (Raissi et al., 2019), which were generated using high-accuracy spectral methods and made publicly available alongside the original PINN paper. These solutions are loaded as spatiotemporal grids and used as ground truth for validation and error analysis. All data are converted to double-precision TensorFlow tensors (float64) to maintain numerical accuracy during training.

Training points are generated using Latin Hypercube Sampling (LHS) to provide well-distributed space-filling coverage of the computational domain. Three training sets are employed for each benchmark: (i) 20,000 interior collocation points to enforce the PDE residuals, (ii) 400 initial-condition points to satisfy temporal constraints, and (iii) 200 boundary condition points to enforce spatial boundary conditions. All training sets are sampled once and held fixed throughout training. Input coordinates are normalized using the bounds of the collocation points to improve numerical stability during training.

### *4.2.2 Training procedure and loss function*

The HQPINN is trained using the composite physics-informed loss introduced in Section 3.5. Training is carried out using a two-stage optimization strategy that is applied consistently across all benchmark problems. In the first stage, the Adam optimizer is applied with a learning rate of 0.001 to enable effective exploration of the parameter space and achieve rapid initial convergence. In the second stage, the limited-memory Broyden-Fletcher-Goldfarb-Shanno (L-BFGS-B) optimizer (Liu & Nocedal, 1989) is employed to refine the solution and improve final accuracy (Berger et al., 2025; Raissi et al., 2019; Urbán et al., 2025). No problem-specific tuning of optimizer hyperparameters is introduced. All models share the same optimizer schedule to ensure fair and controlled comparison across PDEs and sensitivity-analysis cases. Further discussion of the impact of the optimizer is provided in Section 6.2.

### *4.2.3 Performance evaluation protocol*

Model performance is quantified using the relative $L^2$ error over the full spatiotemporal domain as $L^2 = \|u_{\text{pred}} - u_{\text{ref}}\|_2 / \|u_{\text{ref}}\|_2$, where $u_{pred}$ and $u_{ref}$ denote the predicted and reference solutions, respectively. In addition, training dynamics are continuously monitored by real-time tracking the evolution of the individual loss components together with the total physics-informed loss during optimization (Raissi et al., 2019; Yang & Foster, 2022).

### *4.2.4 PDE-specific implementation details*

All three benchmark problems are solved on the spatial domain $x \in [-1, 1]$ and temporal domain $t \in [0, 1]$, with problem-specific initial and boundary conditions. For Burgers' equation, the initial condition is $u(x, 0) = -\sin(\pi x)$ with homogeneous Dirichlet boundary conditions $u(-1, t) = 0$ and $u(1, t) = 0$. For the Allen-Cahn equation, the initial condition is $\phi(x, 0) = x^2\cos(\pi x)$ with non-homogeneous Dirichlet boundary conditions $\phi(-1, t) = -1$ and $\phi(1, t) = -1$. For the KdV equation, the initial condition is $\psi(x, 0) = \cos(\pi x)$, with periodic boundary conditions $\psi(x+L,t)=\psi(x,t)$, where $L = 2$. The grid resolutions, circuit configurations, and iteration schedules used for the reference benchmark cases are summarized in Table 2.

**Table 2.** Numerical parameters and circuit configurations for the reference benchmark cases.

| Equation | Grid resolution | No. of qubits ($n_q$) | No. of variational layers (m) | No. of iterations | |
|---|---|---|---|---|---|
| | | | | Adam | L-BFGS |
| Burgers | 256 × 100 | 5 | 5 | 200 | 300 |
| Allen-Cahn | 512 × 201 | 5 | 5 | 10000 | 15000 |
| KdV | 512 × 201 | 5 | 5 | 5000 | 5000 |

### 4.3 Sensitivity analysis

To examine how HQPINN performance depends on quantum-circuit design, a structured sensitivity analysis is conducted with respect to two key quantum-resource parameters: the number of qubits, $n_q \in \{3,4,5,6,7\}$, and the number of variational layers $m \in \{3,4,5,6,7\}$. These parameters are systematically varied to characterize their impacts on model accuracy, convergence behavior, and training stability across different PDE regimes. All sensitivity experiments employ the same data preprocessing, sampling strategy, optimization procedure, and evaluation metrics described in Section 4.2. Aside from the parameters under investigation, all remaining architectural and training hyperparameters are held fixed. This controlled design ensures that the observed performance variations can be attributed directly to changes in quantum-circuit capacity rather than to confounding effects from the classical network architecture or training procedure.

## 5 Results: training dynamics across PDEs

The results compare the proposed HQPINN with a classical PINN baseline in terms of training dynamics, final solution accuracy as measured by the relative $L^2$ error, and convergence behavior, quantified by the number of iterations required to reach a given accuracy level.

Fig. 2 summarizes the training dynamics of the HQPINN and the classical PINN for the three benchmark PDEs. Fig. 2a-c show the evolution of the total physics-informed loss, while Fig. 2d-f present the corresponding relative $L^2$ error with respect to the reference solution. Across the three benchmarks, the HQPINN generally exhibits smoother loss trajectories, reduced oscillatory behavior, and lower final $L^2$ errors than the classical PINN. These trends indicate that the hybrid quantum-classical architecture produces a more stable training process, particularly during the second-stage L-BFGS-B refinement.

Because the three benchmark PDEs exhibit different levels of stiffness and spectral complexity, different iteration budgets are used: 500 iterations for Burgers' equation, 25,000 for the Allen–Cahn equation, and 10,000 for the KdV equation. These budgets reflect the intrinsic numerical difficulty and convergence characteristics of each problem. Burgers' equation, despite its nonlinearity, converges quickly; the Allen–Cahn equation is stiff and requires substantially longer training; the KdV equation lies between these extremes, with smooth but highly dispersive solutions. For the Burgers' equation (Fig. 2a,d), the HQPINN reaches a final relative $L^2$ error of $2.12 \times 10^{-2}$, compared with $8.46 \times 10^{-2}$ for the classical PINN after 500 iterations. While the two models behave similarly during the Adam phase, the HQPINN shows a much sharper error reduction during the L-BFGS-B stage. Since Burgers' equation develops steep gradients and near-shock features, which standard PINNs often struggle to capture due to spectral bias and the slow learning of high-frequency components, the results suggest that the hybrid model resolves sharp nonlinear structures more effectively than the classical PINN. For the Allen–Cahn equation (Fig. 2b,e), the HQPINN achieves a final relative $L^2$ error of $1.35 \times 10^{-2}$, whereas the classical PINN converges to $7.04 \times 10^{-2}$. The classical PINN shows prolonged oscillations and stagnation during the Adam phase, reflecting the well-known difficulty of training PINNs on stiff reaction-diffusion problems. Such stiffness is known to produce gradients that are back-propagated unequally between the residual loss and the boundary/initial condition loss; it can further cause imbalanced loss functions across state variables when their magnitudes span several orders of magnitude (Ji et al., 2021). In contrast, the HQPINN exhibits a smoother and more sustained decline in both loss and error, indicating improved robustness in this stiff nonlinear regime. For the KdV equation (Fig. 2c,f), both models achieve comparable final accuracy, with the HQPINN reaching $2.07 \times 10^{-2}$ and the classical PINN reaching $2.85 \times 10^{-2}$ after 10,000 iterations. The HQPINN still shows a somewhat smoother optimization trajectory and a modest reduction in final error, but the improvement is less pronounced than for the Burgers' and Allen–Cahn equations. This is consistent with the smoother dispersive character of the KdV solution, for which classical PINNs already provide a reasonably accurate representation. As a result, the hybrid architecture offers only a moderate advantage for the KdV equation, reinforcing the observation that HQPINNs yield the largest gains for PDEs with stronger stiffness, sharp gradients, or more challenging multiscale behavior.

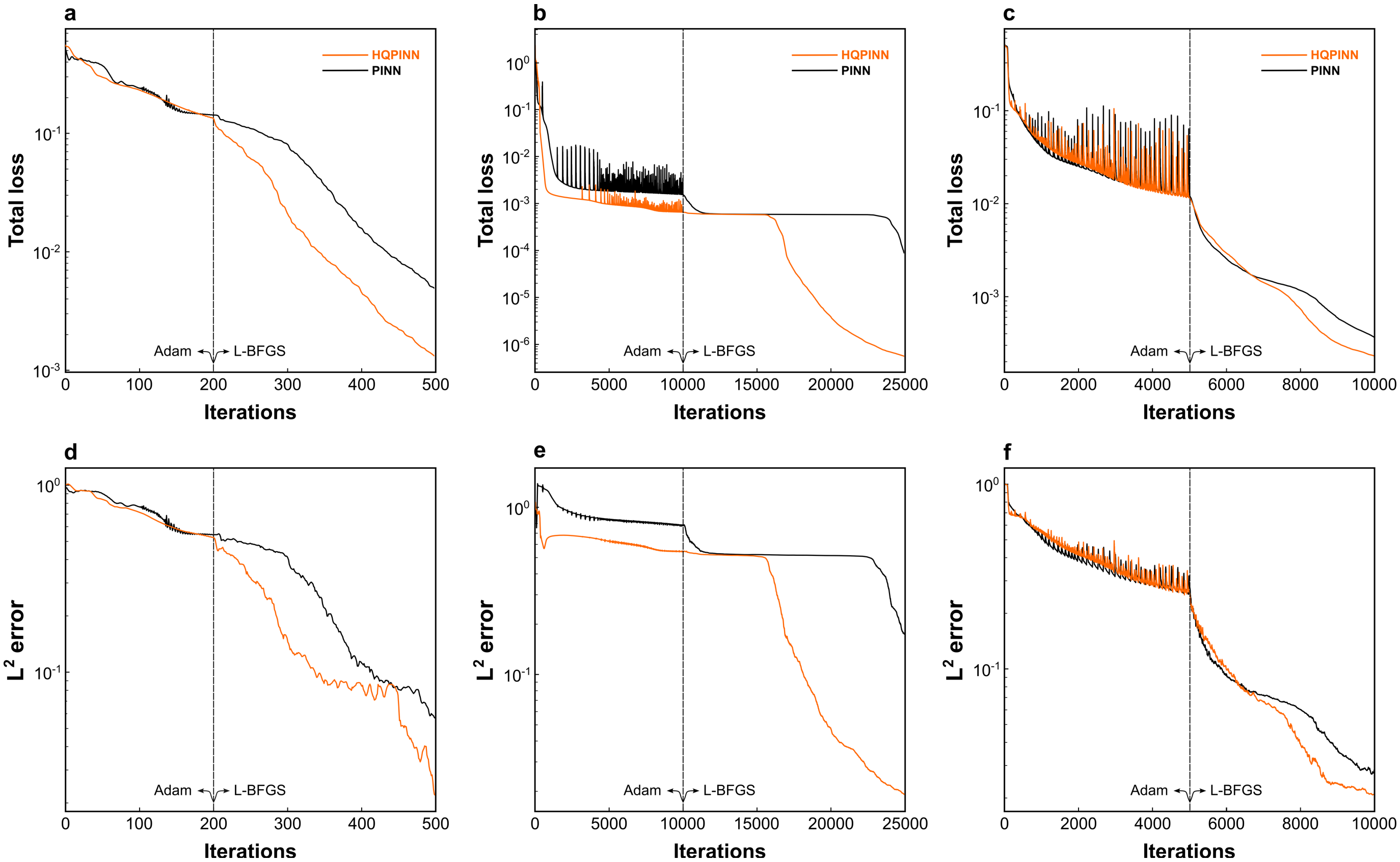

**Fig. 2** Training dynamics of the HQPINN (orange) and the classical PINN (black) for three benchmark PDEs. Total loss during training for (a) Burgers' equation, (b) Allen–Cahn equation, and (c) KdV equation. (d-f) The corresponding relative $L^2$ error with respect to the reference solution. Dashed vertical lines mark the transition from Adam to L-BFGS-B. Training is carried out for 500, 25,000, and 10,000 iterations for Burgers', Allen–Cahn, and KdV equations, respectively.

Overall, Fig. 2 demonstrates that the HQPINN improves training stability and convergence behavior compared with the classical PINN. The most significant gains are observed for the Burgers' and Allen–Cahn equations, which are characterized by sharp gradients, stiffness and pronounced multiscale behavior. At the same time, the hybrid model retains competitive performance for smoother problems such as the KdV equation, indicating that its benefits are problem dependent rather than uniformly dramatic across all PDE types.

Fig. 3 presents the spatial–temporal distribution of the prediction error for the classical PINN and the HQPINN across the three benchmark PDEs. These contour maps complement the global metrics in Fig. 2 by showing where errors are concentrated within the computational domain and how effectively the HQPINN suppresses localized error structures. For Burgers' equation (Fig. 3a,d), the dominant errors in the classical PINN are concentrated near the steep transition region where the solution develops a sharp convective gradient. This appears as a pronounced error band aligned with the shock trajectory. In contrast, the HQPINN substantially reduces this localized error: the shock-aligned structure becomes both narrower and weaker in amplitude. This indicates that the hybrid model captures the steep spatial transition more accurately and confines the error more effectively to the immediate vicinity of the gradient. For the Allen–Cahn equation (Fig. 3b,e), the classical PINN exhibits a strong, vertically oriented error band centered on the metastable interface, consistent with the difficulty of resolving stiff reaction–diffusion dynamics in a narrow interfacial region. The HQPINN, by contrast, markedly suppresses this interface-localized error, leaving only a weaker residual structure. This demonstrates enhanced local accuracy in regions where the solution varies most rapidly, consistent with improved handling of the multiscale gradients characteristic of the Allen–Cahn dynamics. For the KdV equation (Fig. 3c,f), both models display oscillatory error bands that follow the underlying dispersive wave structure. The HQPINN reduces the amplitude of these oscillations, but the improvement is more moderate than in the other two benchmarks. This behavior aligns with the smoother dispersive nature of KdV solutions, which classical PINNs already capture reasonably well. Overall, Fig. 3 shows that the advantages of the HQPINN extend beyond improved global accuracy to enhanced local resolution of steep gradients and sharp interfaces. This spatial–temporal improvement is consistent with the smoother and more stable optimization behavior observed in Fig. 2, and with the increased robustness of the hybrid model across the three benchmark PDEs.

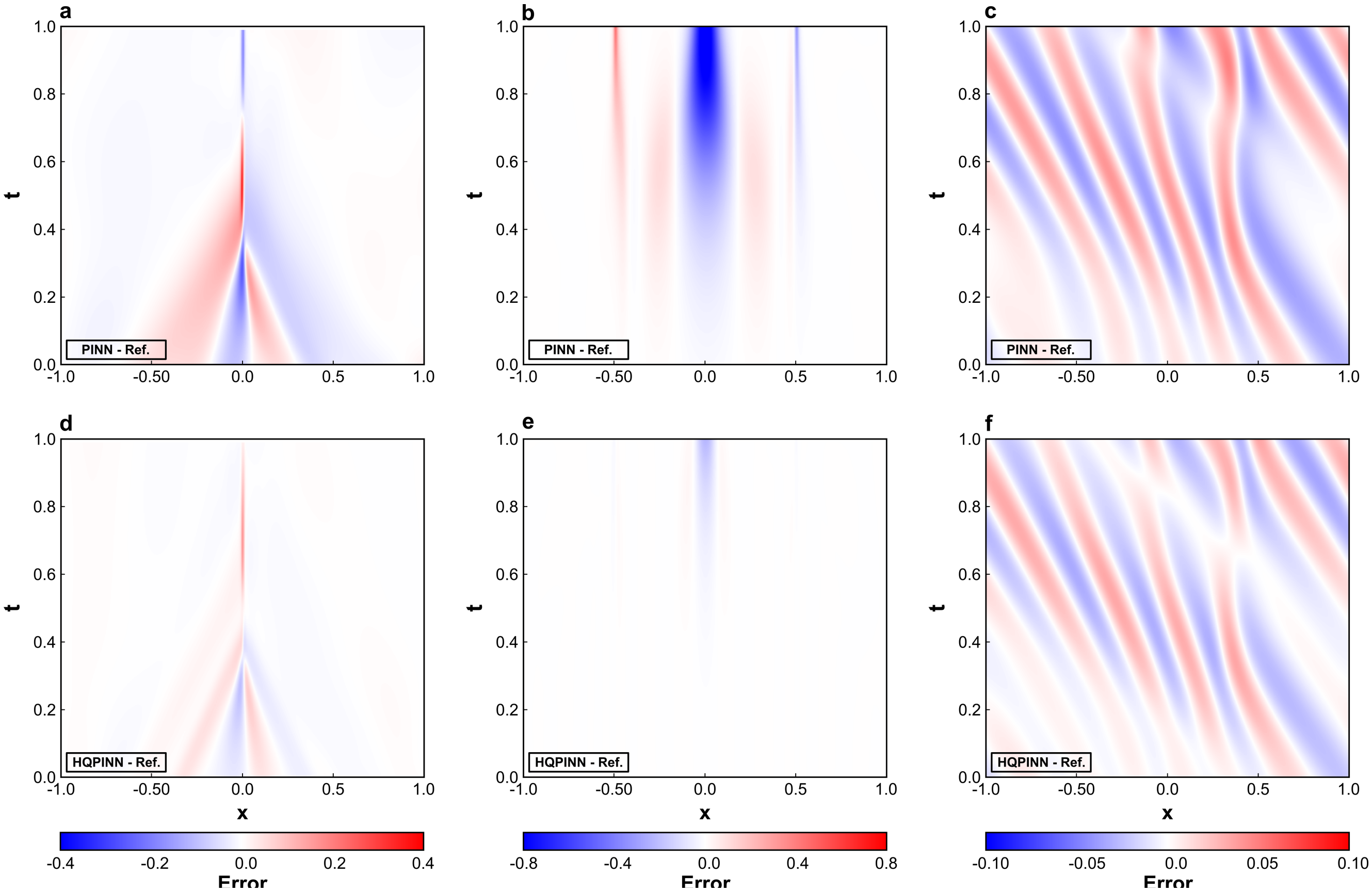

**Fig. 3** Spatial–temporal error fields of the classical PINN for (a) Burgers' equation, (b) Allen–Cahn equation, and (c) KdV equation. (d)–(f) The corresponding error fields of the HQPINN. The plotted quantity is the signed prediction error, defined as the difference between the predicted and reference solutions. For Burgers' and Allen–Cahn equations, the HQPINN substantially reduces both the magnitude and the spatial extent of the dominant localized error regions. For the KdV equation, both models exhibit oscillatory error patterns aligned with the dispersive wave structure, although the HQPINN yields a more modest reduction in error amplitude.

Fig. 4 compares representative solution snapshots of the classical PINN and the HQPINN against the reference solution for the three benchmark PDEs. These profiles complement the spatial-temporal error maps in Fig. 3 by illustrating how accurately each model reproduces key spatial features of the solution at different times. For Burgers' equation, both models closely match the reference solution at early times when the solution remains smooth. As the steep convective gradient forms, the classical PINN begins to deviate in the vicinity of the shock region, whereas the HQPINN remains closely aligned with the reference solution. This behavior reflects improved local resolution of the sharp spatial transition by the hybrid model. A similarly strong contrast is observed for the Allen–Cahn equation (Fig. 4b,e), where the HQPINN more accurately captures the narrow interface structure and the depth of the central metastable region, particularly at later times. For the KdV equation, both models remain close to the reference solution, and the improvement provided by the HQPINN is more moderate, consistent with the smoother dispersive character of the solution. Overall, the solution snapshots show that the advantages of the HQPINN extend beyond reduced global error measures to include improved local agreement in the most challenging regions of the solution domain. Together with the training dynamics in Fig. 2 and the spatial–temporal error distributions in Fig. 3, these results confirm that the hybrid quantum-classical architecture provides its greatest benefits for PDEs characterized by steep gradients or sharply localized features, while maintaining strong performance on smoother problems.

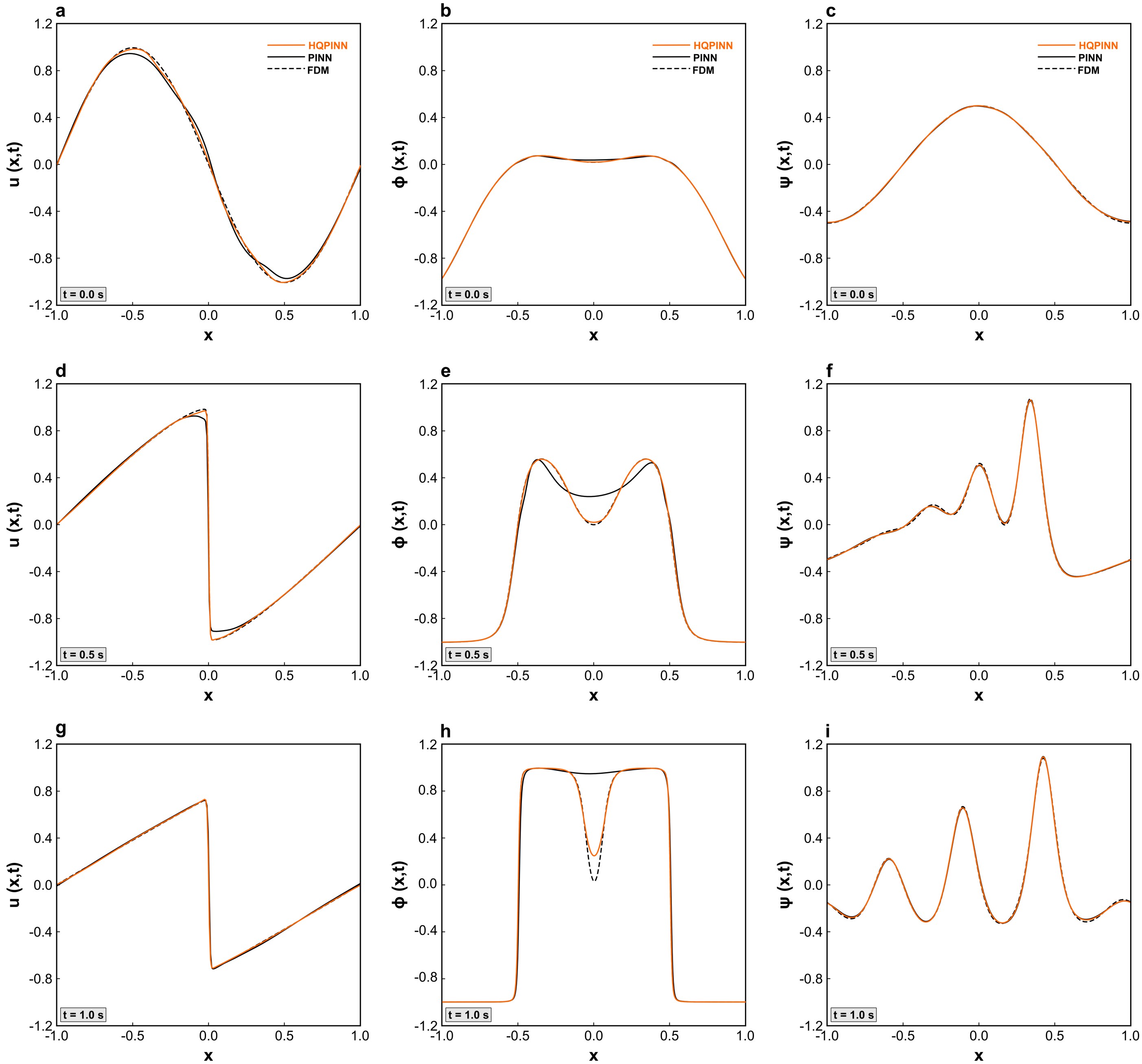

**Fig. 4** Representative solution snapshots of the HQPINN and the classical PINN compared with the reference solution at t = 0 s for (a) Burgers' equation, (b) Allen–Cahn equation, and (c) KdV equation. (d–f) The same comparison at $t$ = 0.5 s, and (g–i) at t = 1.0 s. For the Burgers' and Allen–Cahn equations, the HQPINN more accurately reproduces the steep gradient and narrow interfacial structures, respectively, while for the KdV equation both models capture the dispersive wave profile well, with a more modest advantage for the HQPINN. The orange curves represent the HQPINN predictions, the black lines correspond to the conventional PINN results, and the dashed lines indicate the reference solutions.

## 6 Results: sensitivity analysis

### 6.1 Impact of number of qubits and quantum layers

Fig. 5 shows how the relative $L^2$ error varies as a function of the number of qubits and the number of variational layers for the three benchmark PDEs. In this parametric study, both quantum-resource parameters are varied from 3 to 7, while the classical backbone is kept fixed at 6 hidden layers with 40 neurons per layer. To ensure comparable qubit-layer scans within each benchmark, the iteration budgets are set to 5,000 for Burgers' equation, 20,000 for the Allen–Cahn equation, and 10,000 for the KdV equation. Across all three problems, the heat maps reveal a clear non-monotonic dependence on both qubit count and circuit depth. Increasing the number of qubits or variational layers does not guarantee improved accuracy; instead, each PDE exhibits a distinct optimal region in the qubit–layer design space beyond which performance plateaus or degrades. This indicates that the best-performing quantum-circuit configuration is strongly problem dependent under the adapted training protocol.

For Burgers' equation (Fig. 5a), the minimum error, $L^2 = 2.54 \times 10^{-4}$, is obtained with 5 qubits and 3 variational layers. The best performance is therefore achieved with a relatively shallow quantum circuit, and increasing circuit depth or qubit count beyond this configuration does not yield further improvement within the allocated iteration budget, which is consistent with previous observations that smooth or mildly nonlinear PDEs can be modelled effectively with shallow PQCs (Markidis, 2022; Trahan et al., 2024). For the Allen–Cahn equation (Fig. 5b), the optimal configuration shifts toward larger circuits, with the minimum error occurring at 6 qubits and 7 variational layers ($L^2 = 3.13 \times 10^{-2}$). This result suggests that deeper quantum circuits are beneficial for stiff reaction–diffusion dynamics up to an optimal depth, beyond which performance begins to degrade. For the KdV equation (Fig. 5c), the lowest error ($L^2 = 1.85 \times 10^{-2}$) is obtained with 7 qubits and 4 variational layers, with performance varying more moderately across the scanned range. Compared with the Allen-Cahn case, these results indicate a weaker sensitivity to circuit depth. This is consistent with a smaller benefit from additional circuit depth for the smoother dispersive structure of the KdV solution, while moderate increases in qubit count still improve performance up to the optimal configuration. Overall, Fig.5 demonstrates that the optimal quantum-circuit configuration in HQPINNs is strongly problem dependent. The non-monotonic trends observed across all three benchmarks show that increasing the number of qubits or variational layers does not inherently lead to improved accuracy. Instead, each PDE admits a distinct optimal region in the qubit–layer design space, beyond which additional circuit capacity yields diminishing returns or degraded performance. These results highlight the need to balance circuit size and trainability when designing effective HQPINN architectures(Farea et al., 2025.; Trahan et al., 2024).

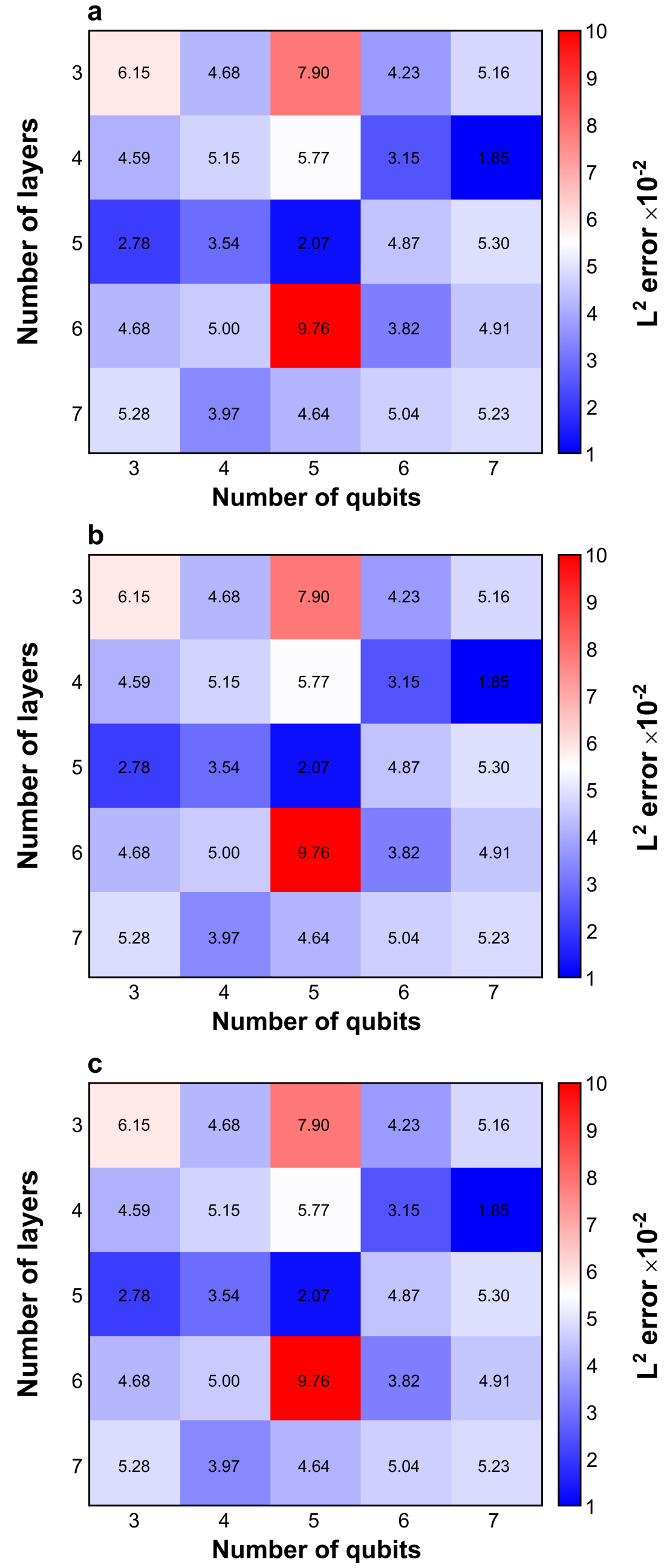

**Fig. 5** Heat maps of the final error of the HQPINN for different quantum-circuit configurations, parameterized by the number of qubits and the number of variational layers, for (a) Burgers' equation, (b) Allen–Cahn equation, and (c) KdV equation. The classical part of the network architecture is kept fixed in all cases. The results show a non-monotonic dependence of accuracy on quantum-resource allocation, with a distinct optimal region for each benchmark PDE.

## 6.2 Impact of PQC position

Fig. 6 shows how the placement of the parameterized quantum circuit (PQC) within the hybrid architecture affects HQPINN performance for the three equations. Three configurations are compared: (i) PQC placement before the first classical hidden layer, directly after the input, (ii) PQC placement in the middle of the classical hidden layers, and (iii) PQC placement after the final classical hidden layer, before the output layer. All models are trained using the same two-stage Adam + L-BFGS optimization procedure (see Table 2). Across all three PDEs, placing the PQC after the final classical layer yields the lowest final error and the most stable convergence trajectory. Input-stage placement leads to slower convergence and substantially higher errors, while mid-network placement produces intermediate performance. The advantage of output-stage placement becomes most pronounced after the transition from Adam to L-BFGS, particularly for the Burgers' and KdV equations, where this configuration benefits most strongly from second-stage refinement. Because the optimization schedule and iteration budgets are identical across all configurations, these results indicate that output-stage placement is the best-performing choice within the present training protocol.

A plausible interpretation is that placing the PQC near the output allows the quantum layer to act on a latent representation that has already been shaped by multiple nonlinear classical transformations. In this configuration, the quantum layer operates on more structured intermediate features than would be available at the input stage, which may be advantageous under the fixed qubit count and circuit depth used in this study. This interpretation is consistent with prior hybrid quantum–classical learning results suggesting that variational quantum circuits are often most effective when applied to latent features rather than directly to raw inputs, particularly in regression-oriented settings (Long et al., 2025; Trahan et al., 2024). By contrast, early quantum embeddings, where quantum circuits act directly on raw inputs, can be more advantageous in classification or kernel methods, where the quantum feature map is intended to dominate the representational power (Havlíček et al., 2019).

Overall, Fig. 6 highlights that PQC placement is a critical architectural design choice in HQPINNs. Within the present framework, positioning the PQC near the output provides the most favorable combination of convergence behavior and final accuracy across all three benchmark PDEs.

### 6.3 Impact of number of training points

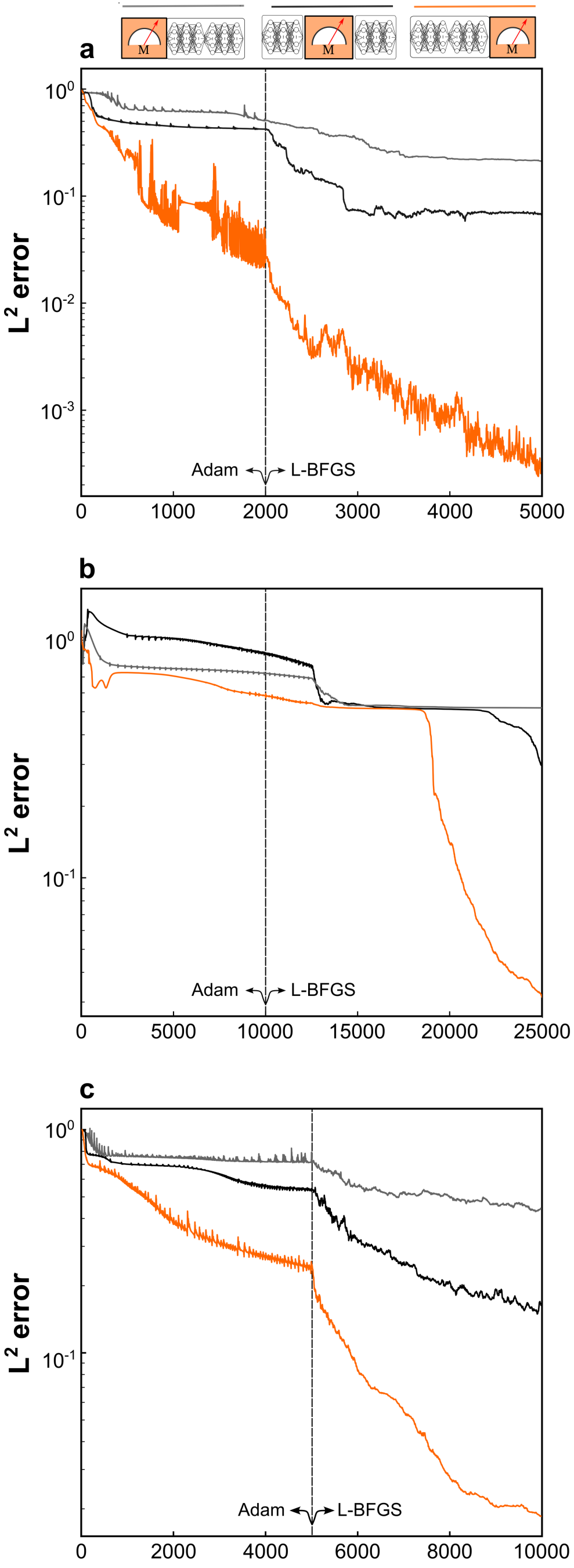

**Fig. 6.** Sensitivity of HQPINN performance to the position of the parameterized quantum circuit for (a) Burgers' equation, (b) Allen–Cahn equation, and (c) KdV equation. The relative $L^2$ error s plotted against training iterations for three PQC placements: input stage, middle hidden-layer stage, and after the final classical hidden layer. Dashed vertical lines mark the transition from Adam to L-BFGS-B. The output-stage PQC consistently achieves the lowest final error and the smoothest convergence.

Fig. 7 shows how the relative $L^2$ error varies with the number of training points for the HQPINN and the classical PINN across the three benchmark PDEs. The total number of training samples is set to 150, 300, 450, and 600 points, while the relative proportions of interior, initial-condition, and boundary-condition points are kept fixed to preserve a consistent balance of physical constraints. This balanced distribution follows standard practices in PINN literature (Wu et al., 2022) and ensures that neither initial/boundary constraints nor interior physics dominate the loss function. For Burgers' equation (Fig. 7a), the HQPINN consistently achieves lower errors than the classical PINN across all sampling levels. The hybrid model is especially robust in the low-data regime (150 points), maintaining stable accuracy even with only 150 samples, whereas the classical PINN deteriorates markedly. This indicates reduced sensitivity of the HQPINN to sparse collocation for this nonlinear transport problem. For the Allen–Cahn equation (Fig. 7b), the advantage of the hybrid model is even more pronounced. The classical PINN exhibits large errors across all sampling densities, reflecting the well-known difficulty of training PINNs on stiff reaction–diffusion equations and narrow metastable interfaces without dense collocation (Wang et al., 2021). In contrast, the HQPINN maintains substantially lower and more stable error levels throughout the full range of training points. This trend is consistent with prior studies showing that Allen–Cahn-type problems are particularly challenging for standard PINNs and often require improved approximation power and more effective sampling strategies (Wang et al., 2021; Wight & Zhao, 2020). For the KdV equation (Fig. 7c), a different trend is observed. At the lowest sampling level (150 points), the HQPINNs perform worse than the classical PINN. However, once the number of training points reaches 300 or more, the HQPINN consistently outperforms the classical model, with the performance gap increasing at 450 and 600 points. This suggests that, for dispersive PDEs involving higher-order derivatives, the hybrid model requires a sufficiently dense set of physical constraints before its advantage becomes apparent. Overall, Fig. 7 shows that the sensitivity of HQPINNs to the number of training points is strongly problem dependent. One possible interpretation of these trends is that parameterized quantum circuits introduce richer nonlinear feature transformations of the encoded inputs, enabling the hybrid model to exploit limited physical constraints more effectively. This view is consistent with prior results showing that data-encoded PQCs can realize Fourier-like function classes, while PINN performance remains strongly influenced by the collocation strategy (Mandl et al., 2023; Sharma et al., 2022).

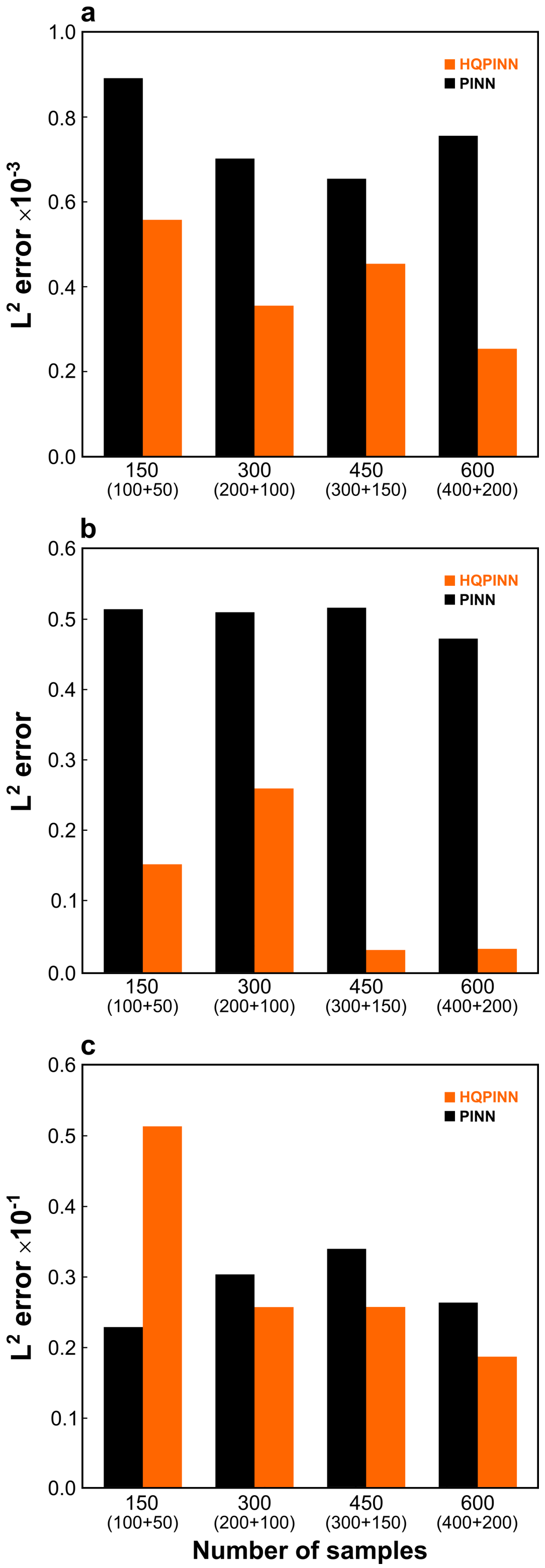

**Fig. 7** The sensitivity of HQPINN and classical PINN accuracy to the number of training samples for (a) Burgers', (b) Allen–Cahn, and (c) KdV equations. Bars show the final relative $L^2$ error obtained using 150, 300, 450, and 600 training points. The HQPINN consistently outperforms the classical PINN for Burgers' and Allen–Cahn equations across all sampling levels, whereas for the KdV equation, the hybrid model requires a higher sampling density before achieving a lower error than the classical PINN.

### 6.4 Impact of network width

Fig. 8 shows the sensitivity of the relative $L^2$ error to the width of the classical network for both the HQPINN and the classical PINN across the three benchmark PDEs. Here, the network width, defined as the number of neurons per hidden layer, is varied over 10, 20, 30, and 40, while the network depth, collocation points, optimizer settings, and all other hyperparameters are kept fixed. The resulting trends illustrate how effectively each architecture benefits from increased classical-network capacity. Across all three PDEs, the HQPINN exhibits a more pronounced and consistent reduction in error with increasing width than the classical PINN. For Burgers' equation (Fig. 8a), both models show improved accuracy as the number of neurons increases; however, the HQPINN achieves lower error at every tested width. The difference becomes particularly notable at 40 neurons, where the HQPINN reaches approximately $0.025 \times 10^{-3}$ compared to $0.075 \times 10^{-3}$ for the classical PINN, corresponding to roughly a threefold reduction in error. For the Allen–Cahn equation (Fig. 8b), the contrast is much stronger. The classical PINN remains near a high error level for widths of 10, 20, and 30 neurons, and exhibits only marginal improvement at 40 neurons. In contrast, the HQPINN exhibits a steep and sustained reduction in error, decreasing from approximately 0.51 at width 10 neurons to around 0.03 at 40 neurons. This represents more than an order-of-magnitude improvement and demonstrates that the hybrid model is considerably more effective at exploiting additional classical-network capacity for this stiff reaction-diffusion benchmark. For the KdV equation (Fig. 8c), both models benefit from increased width, and the HQPINN consistently maintains lower error than the classical PINN across all tested configurations. However, the performance gap is smaller than for the Allen–Cahn benchmark, indicating that the benefit of widening the classical backbone is more moderate for this smoother, dispersive problem. Taken together, these results show that increasing network width benefits the HQPINN more consistently than the classical PINN, particularly for problems characterized by stiffness or sharp spatial features. A plausible interpretation is that a wider classical backbone provides a richer latent representation before the quantum circuit is applied, allowing the hybrid model to make more effective use of the downstream quantum transformation (Cerezo et al., 2021; Wang et al., 2021).

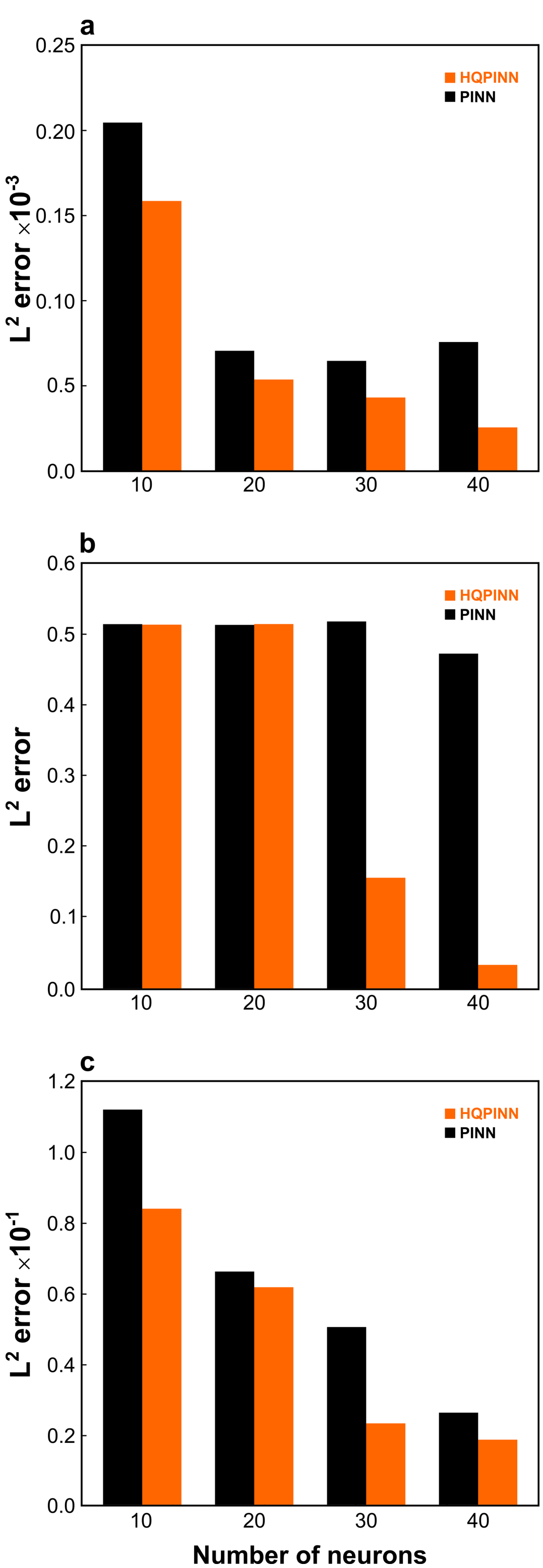

**Fig. 8** Sensitivity of HQPINN and classical PINN accuracy to the number of neurons per hidden layer for (a) Burgers', (b) Allen–Cahn, and (c) KdV equations. Bars show the final relative $L^2$ error for network widths of 10, 20, 30, and 40 neurons per layer. The HQPINN achieves lower error than the classical PINN across all tested widths, with the most pronounced advantage observed for the Allen–Cahn equation.

## 7 Discussion

### 7.1 Mechanisms underlying hybrid quantum–classical gains

The results suggest that the performance advantage of HQPINNs over classical PINNs does not arise from any single architectural component in isolation, but from the interaction between the classical backbone, the parameterized quantum circuit, and the physics-informed objective. In the proposed architecture, the classical network first maps the spatiotemporal coordinates onto a structured latent representation, after which the PQC acts on this representation through data encoding, variational transformation, and expectation-value measurement. One useful interpretive perspective, consistent with prior theoretical results on data-encoded PQCs, is that such circuits can induce finite trigonometric function classes whose accessible frequency content depends on the encoding strategy and circuit structure (Cerezo et al., 2021). Under this view, the PQC may provide a compact mechanism for introducing oscillatory structure into the latent representation. Classical neural networks, including standard PINN backbones, are known to exhibit spectral bias, meaning that low-frequency content is typically learned more readily than a high-frequency or sharply localized structure. In PINNs, this challenge is further compounded by the multi-term physics-informed objective, where PDE- residual, initial-condition, and boundary-condition terms may generate strongly imbalanced gradients and difficult optimization dynamics. (Wang et al., 2021). From this standpoint, the hybrid architecture can be viewed as providing an additional representational mechanism that may help the model handle more effectively once informative latent features have been formed. This interpretation is broadly consistent with earlier quantum-enhanced PINN studies reporting competitive or improved accuracy on selected PDEs, particularly for oscillatory or structured solutions (Kyriienko et al., 2021; Yew Leong et al., 2025), although the present study does not directly identify the precise mechanism responsible for the observed gains.

Beyond representational effects, the results also suggest that the hybrid architecture may influence optimization behavior. In the present results, HQPINNs generally exhibit smoother training trajectories, reduced loss oscillations, and more stable convergence (e.g. Fig. 2), particularly during the L-BFGS stage. This optimizer-dependent behavior is informative because the Adam phase mainly performs a broad first-order exploration of parameter space, whereas L-BFGS acts as a local quasi-Newton refinement and is therefore more sensitive to the conditioning of the learned representation. The stronger performance of HQPINNs during this second stage is consistent with the possibility that the hybrid model provides a more favorable local optimization landscape once the main solution manifold has been approximately identified. However, this interpretation should remain cautious. The present study does not directly analyze Hessian structure, gradient statistics, or optimizer dynamics, and variational quantum circuits are not universally easier to train. The current results therefore support an empirical optimization advantage, but not yet a definitive mechanistic explanation. Taken together, the findings suggest that HQPINNs are most beneficial when the target PDE places simultaneous demands on representation and optimization, particularly in the presence of stiffness, sharp gradients, or multiscale structure. In smoother regimes, such as the KdV benchmark considered here, the hybrid advantage is more modest, indicating that the benefit of the quantum component is problem dependent rather than universal.

### 7.2 Limitations and outlook

The following points concern the scope and design of the present study:

- The present study is based solely on simulations, with all experiments performed using classical quantum-circuit simulators under effectively noiseless conditions. As a result, the reported gains do not account for hardware-level effects such as gate noise, decoherence, limited qubit connectivity, shot noise in expectation-value estimation, or measurement overhead, all of which can materially affect trainability and attainable accuracy, especially as qubit counts and circuit depths increase (Preskill, 2018; Cerezo et al., 2021). This issue is particularly important because the variational-quantum literature has shown that trainability is not guaranteed even in idealized settings: variational circuits may suffer from barren plateaus, and hardware noise can further suppress gradients as system size or circuit depth grows (McClean et al., 2018). The present results should therefore be interpreted as evidence of algorithmic potential under idealized simulation, rather than as a direct predictor of near-term hardware performance. Establishing whether similar improvements persist under realistic quantum noise models or on physical devices will require targeted hardware validation or, at a minimum, systematic benchmarking under calibrated noise channels.
- A second limitation is ansatz dependence of the present conclusions. The study considers one PQC family, one encoding strategy, and a finite set of classical–quantum design variations. Although the sensitivity analysis explores qubit count, variational depth, PQC placement, sampling density, and classical-network width, the space of possible hybrid architectures is much broader. Different encoding schemes, entangling structures, measurement observables, or data-reuploading strategies could lead to different performance trends. The present results should therefore be interpreted as specific to the HQPINN design studied here, rather than as a universal assessment of all hybrid quantum–classical PINN formulations.
- The reported optima are conditional on the training protocol adopted in this work. All comparisons are conducted under fixed optimizer settings, finite iteration budgets, and a common two-stage Adam + L-BFGS training procedure. While this controlled setup is essential for a fair comparison, it also means that the best-performing qubit counts, circuit depths, and widths reported here should be interpreted as conditional optima under the present computational budget, not as absolute optima under unlimited training or exhaustive hyperparameter search. Deeper or wider circuits may perform differently under alternative optimizer schedules, initialization strategies, or longer training horizons.
- Although the present study examines the effect of training-point density, the broader collocation strategy is not systematically varied. In particular, the spatial and temporal placement of collocation points, the balance between residual, initial-condition, and boundary-condition samples, and adaptive refinement strategies are kept fixed. Since PINN performance is known to depend strongly on collocation design, especially in regions of rapid solution variation, the trends reported here should be interpreted under the sampling strategy adopted in this study rather than as a comprehensive evaluation of HQPINN sensitivity to collocation design in general.
- The scope of the benchmark set is limited to three one-dimensional canonical PDEs, which provide a controlled setting for isolating quantum–classical interactions but do not capture the full complexity of higher-dimensional, multiphysics, or engineering-scale systems. Whether the trends observed here extend to more advanced PDE settings, therefore remains an open question. Future work should test HQPINNs on higher-dimensional PDEs, coupled multiphysics systems, and application-driven engineering problems, where increased input dimensionality, stronger nonlinearity, and more complex boundary conditions may alter the balance between quantum-circuit capacity, classical-network design, and training stability.

- Finally, the present study does not provide a full wall-clock or complexity analysis. Although the results show that HQPINNs can improve convergence behavior and final accuracy under the tested conditions, this should not be interpreted as evidence of general runtime speedup. The cost of circuit simulation and differentiation remains substantial, and a more complete computational analysis would be required before stronger claims about efficiency could be made.

Taken together, these limitations indicate that the present work should be viewed as a controlled benchmark study of HQPINN behavior under a specific set of architectural, optimization, and simulation assumptions. Within that scope, the results are informative and encouraging, but broader conclusions will require noise-aware benchmarking, wider architectural exploration, more systematic comparison protocols, and extension to higher-dimensional problems.

## 8 Conclusions

This study developed and systematically evaluated a hybrid quantum-classical physics-informed neural network (HQPINN) framework for solving nonlinear partial differential equations. By integrating a classical neural-network backbone with a parameterized quantum circuit within a physics-informed learning paradigm, the proposed approach was benchmarked against a classical PINN on three representative one-dimensional nonlinear PDEs: Burgers', Allen–Cahn, and Korteweg–de Vries (KdV) equations. In addition to the baseline comparison, an extensive sensitivity analysis is performed to assess how the quantum-circuit configuration, PQC placement, sampling density, and classical-network width influence hybrid model behavior.

The main conclusion of this study is that HQPINNs can achieve more stable training dynamics and improved solution accuracy than classical PINNs, but that this advantage is strongly problem-dependent. Across the benchmark set, the hybrid model generally exhibited smoother optimization trajectories, reduced loss oscillations, and lower final relative $L^2$ errors. The most pronounced gains are observed for the Burgers' and Allen–Cahn equations, where sharp gradients, stiffness, and localized structures pose substantial challenges for classical PINNs. For the smoother and more dispersive KdV equation, the hybrid advantage remained present but was more modest, indicating that HQPINNs are most beneficial in regimes that impose stronger representational and optimization demands.

The sensitivity analysis yields several practical design insights:

- Optimal quantum-circuit configurations are not universal: the best combination of qubit count and variational depth varies across PDE types, and increasing quantum resources does not automatically improve performance.
- PQC placement plays a critical role: placing the quantum circuit after the final classical layer consistently produces the best convergence behavior and accuracy, highlighting the importance of operating on a structured latent representation rather than on raw inputs.
- Increasing classical-network width benefits HQPINNs more consistently than classical PINNs, underscoring the continued importance of classical feature extraction within hybrid architectures.
- The effect of training-point density is also problem dependent: HQPINNs show greater robustness to sparse sampling for Burgers' and Allen–Cahn equations, whereas for the KdV equation their advantage emerges only once the collocation density becomes sufficiently high.

Overall, the present results demonstrate that carefully constructed hybrid quantum–classical architectures can mitigate important limitations of classical PINNs in challenging nonlinear regimes, particularly those involving stiffness, sharp gradients, or multiscale structure.

## Author contributions

**KZ:** Methodology, Conceptualization, Software, Validation, Formal analysis, Investigation, Visualization, Writing – original draft, Writing – review & editing. **HM:** Methodology, Supervision, Conceptualization, Visualization, Writing – original draft, Writing – review & editing. **AS:** Methodology, Supervision, Writing – review & editing.